\documentclass[12pt,preprint]{aastex}

\shorttitle{Supernova SN 2005bf}
\shortauthors{Anupama et al.}
\begin{document}

\title{The Peculiar Type Ib Supernova SN 2005bf: Explosion of a Massive He Star
With a Thin Hydrogen Envelope?}
\author{G. C. Anupama\footnote{gca@iiap.res.in}, D. K. Sahu\footnote{dks@crest.e
rnet.in},}
\affil{Indian Institute of Astrophysics, Koramangala, Bangalore 560 034, India}
\author{J. Deng\footnote{jsdeng@bao.ac.cn},}
\affil{National Astronomical Observatories, CAS, 20A Datun Road,
Chaoyang District, Beijing 100012, China}
\author{K. Nomoto\footnote{nomoto@astron.s.u-tokyo.ac.jp}, N. Tominaga\footnote{
tominaga@astron.s.u-tokyo.ac.jp}, M. Tanaka\footnote{mtanaka@astron.s.u-tokyo.ac
.jp},}
\affil{Department of Astronomy, University of Tokyo, Hongo 7-3-1, Bunkyo-ku, Tok
yo 113-0033, Japan}
\author{P. A. Mazzali\footnote{mazzali@MPA-Garching.MPG.DE,}}
\affil{Department of Astronomy, University of Tokyo, Hongo 7-3-1, Bunkyo-ku, Tok
yo 113-0033, Japan.}
\affil{Max-Planck Inst. f\"ur Astrophysik, Garching, Germany}
\affil{INAF - Oss. Astron. Trieste, Italy}
\author{T. P. Prabhu\footnote{tpp@iiap.res.in}}
\affil{Indian Institute of Astrophysics, Koramangala, Bangalore 560 034, India}
\newpage
\begin{abstract}

We present $BVRI$ photometry and optical spectroscopy of SN 2005bf near light
maximum. The maximum phase is broad and occurred around 2005 May 7, about
forty days after the shock breakout. SN 2005bf has a peak bolometric magnitude
$M_{\rm{bol}}=-18.0\pm 0.2$: while this is not particularly bright, it
occurred at an epoch significantly later than other SNe Ibc, indicating that
the SN possibly ejected $\sim 0.31$~M$_\sun$ of $^{56}$Ni, which is more than
the typical amount. The spectra of SN 2005bf around maximum are very similar
to those of the Type Ib SNe 1999ex and 1984L about 25-35 days after explosion,
displaying prominent He I, Fe II, Ca II H \& K and the near-IR triplet P Cygni
lines. Except for the strongest lines, He I absorptions are blueshifted by
$\lesssim 6500$ km s$^{-1}$, and Fe II by $\sim 7500-8000$ km s$^{-1}$.
No other SNe Ib have been reported to have their Fe II absorptions blueshifted
more than their He I absorptions. Relatively weak H$\alpha$ and very weak 
H$\beta$ may also exist, blueshifted by $\sim 15,000$ km s$^{-1}$. We suggest 
that SN 2005bf was the explosion of a massive He star, possibly with a trace of 
a hydrogen envelope.

\end{abstract}


\section{Introduction}

The study of a subclass of hydrogen deficient supernovae, namely type Ib and
type Ic events, has been one of the interesting topics in supernova (SN)
research. The observational properties, progenitors, and hence the physics of
explosion are the least understood for these two subclasses. The recently
established connection of bright and energetic SN Ic events with Gamma-Ray
Burst sources \citep[e.g.][]{maz03} makes their study interesting and exciting.
Type Ib and Ic SNe are classified on the basis of their spectra. Both types are
hydrogen deficient at maximum light, and also lack the deep Si II absorption
near 6150~\AA, a characteristic feature of the Ia events. At early phases Ic's
do not show lines of He I, shown by the Ib's, while at later phases both types
have similar spectra \citep{wh90, mat01, b02}. The Ib and Ic supernovae events
are widely accepted to be core collapse supernovae \citep[e.g.][]{shi90, hac91,
wos93, nom94}.

Supernova SN 2005bf was discovered independently by \citet{mon05} and
\citet{ml05} on April 5.722, 2005 (UT), at a magnitude about 18.0,
in the SB(r)b galaxy MCG +00-27-5. The supernova was also marginally detected,
at magnitude about 18.8, on a image taken on Mar. 30.31 \citep{ml05}. Early
spectroscopic observations indicated the supernova to be of type Ic a few days
before maximum light \citep{mor05, mod05}.
SN 2005bf was reported to undergo an unusual photometric behaviour by
\citet{ham05}. After an initial brightening from April 7 to April 13, the
supernova declined until April 21, after which it re-brightened to magnitudes
brighter than the initial maximum (see light curves posted at the Carnegie
Supernova Project (CSP) group website{\footnote{http://csp1.lco.cl/$\sim$cspuser1/images/OPTICAL\underbar{ }LIGHT\underbar{ }CURVES/SN05bf.html}}).
Spectra obtained during the re-brightening \citep{wb05, mkc05} indicated that
the spectrum had developed conspicious lines of He I similar to type Ib
supernovae. Also, the SN reached a bright maximum, making it an interesting
target.

In this paper we present optical spectroscopy of SN 2005bf obtained near the 
maximum, and optical photometry during the maximum and subsequent decline. CCD
photometric and spectroscopic observations were performed with the 2-m
Himalayan Chandra Telescope (HCT) at the Indian Astronomical Observatory (IAO),
Hanle, India using the Himalaya Faint Object Spectrograph Camera (HFOSC).

\section{The Optical and Bolometric Light Curves}

\begin{figure}
{\includegraphics[width=18cm]{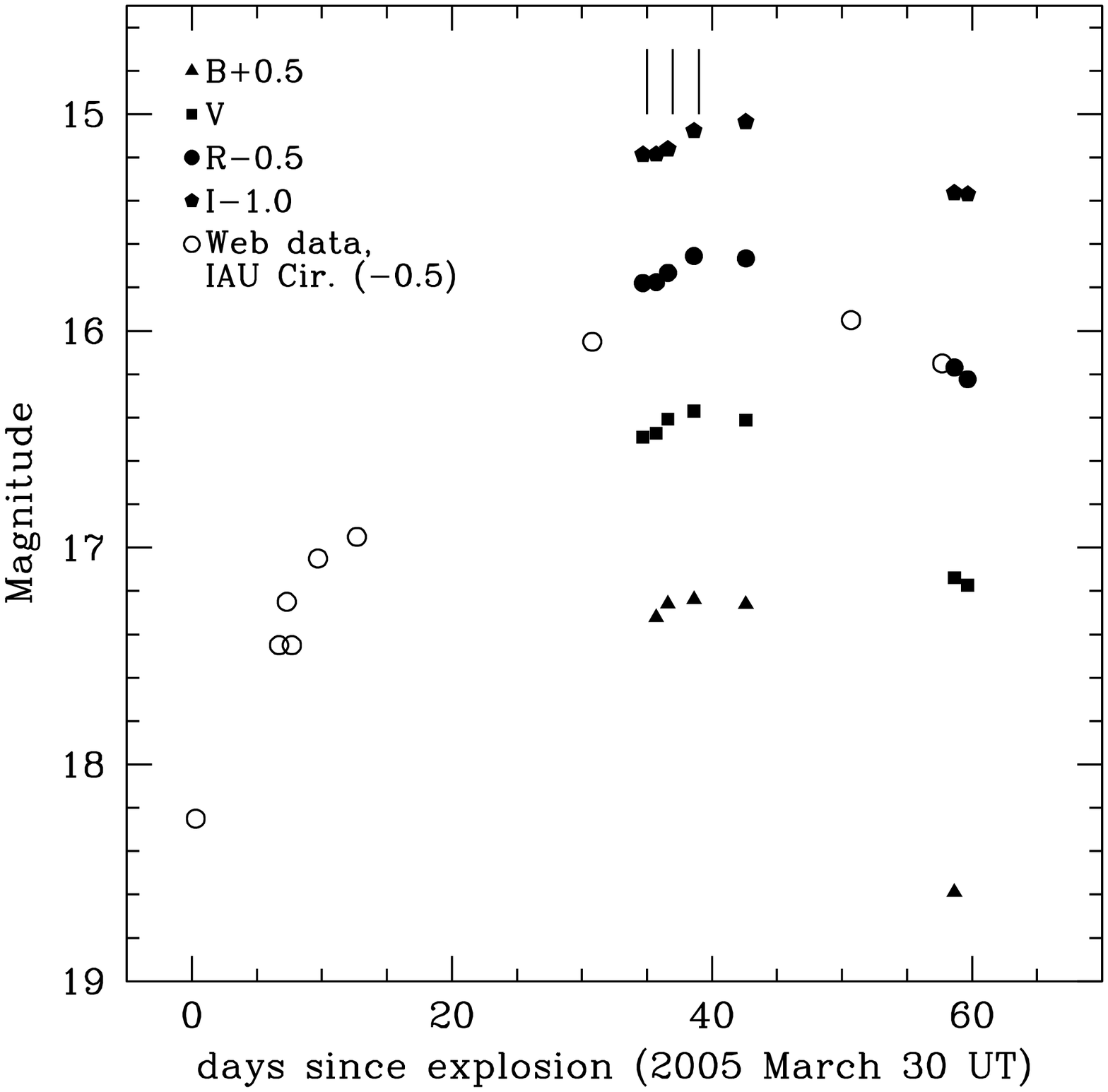}}
\caption{The $BVRI$ magnitudes of SN 2005bf. For clarity the $BRI$ magnitudes
are offset by $+0.5$, $-0.5$ and $-1.0$ magnitudes, respectively. Also included
in the figure with the $R$ band magnitudes are the unfiltered CCD magnitudes
obtained by amateurs and those reported in the IAU circulars. Vertical lines
mark the dates of spectroscopic observations.
\label{fig1}}
\end{figure}

Photometric observations in the Bessell $BVRI$ bands were made during
May 3\,--\,May 28. Landolt standard regions were observed on May 27 to
calibrate a sequence of secondary standards in the supernova field.
The magnitudes of SN 2005bf and the secondary
standards in the field were obtained by point spread function photometry.
The identification and magnitudes of the secondary standards in the field of
SN 2005bf may be obtained from http://www.iiap.res.in/personnel/gca/sn05cal.html.
The $BVRI$ light curves of SN 2005bf are shown in Figure \ref{fig1}. Also
included in the figure with the $R$ magnitudes are the unfiltered CCD
magnitudes reported in the IAU Circulars, and the estimates made by amateurs
{\footnote{http://www.astrosurf.com/snweb2/2005/05bf/05bfMeas.htm}}.
The pre-maximum evolution of SN 2005bf was quite peculiar and different from
that of other SNe Ib/c. SN 2005bf had a very slow rise to the maximum, which
occurred around 2005 May 7, nearly 40 days since the shock breakout. In
contrast, the type Ib SN 1999ex rose to maximum in about 18 days \citep{set02}.
After maximum, which was broad, SN 2005bf declined with rates of 0.07
mag~day$^{-1}$, 0.038 mag~day$^{-1}$, 0.05 mag~day$^{-1}$ and 0.014
mag~day$^{-1}$ in $B$,$V$,$R$ and $I$, respectively, which are similar to the
decline rates observed in SN Ib SN 1999ex at similar epochs with respect to
maximum.

In this paper we assume the date of explosion to be 2005 March 30
(JD\,2\,45\,3459.5), based on \citet{ml05} and the light curve posted by the
CSP. A Galactic reddening E(B-V) = 0.045 as estimated by \citet{scet98} in the
direction of the
host galaxy has been used. As the supernova occurred in one of the spiral arms
of the host galaxy, reddening within the host is also expected. However, since
no conspicious Na I D absorption features have been reported, we assume
negligible extinction due to the host galaxy.  We adopt a distance modulus of
$\mu=34.5$ for the host galaxy using $H_0=72$~km~s$^{-1}$~Mpc$^{-1}$, $\Lambda
= 0.7$, $\Omega_M=0.3$, and a redshift of $z=0.0188$ (HyperLeda database).

\begin{figure}
{\includegraphics[width=18cm]{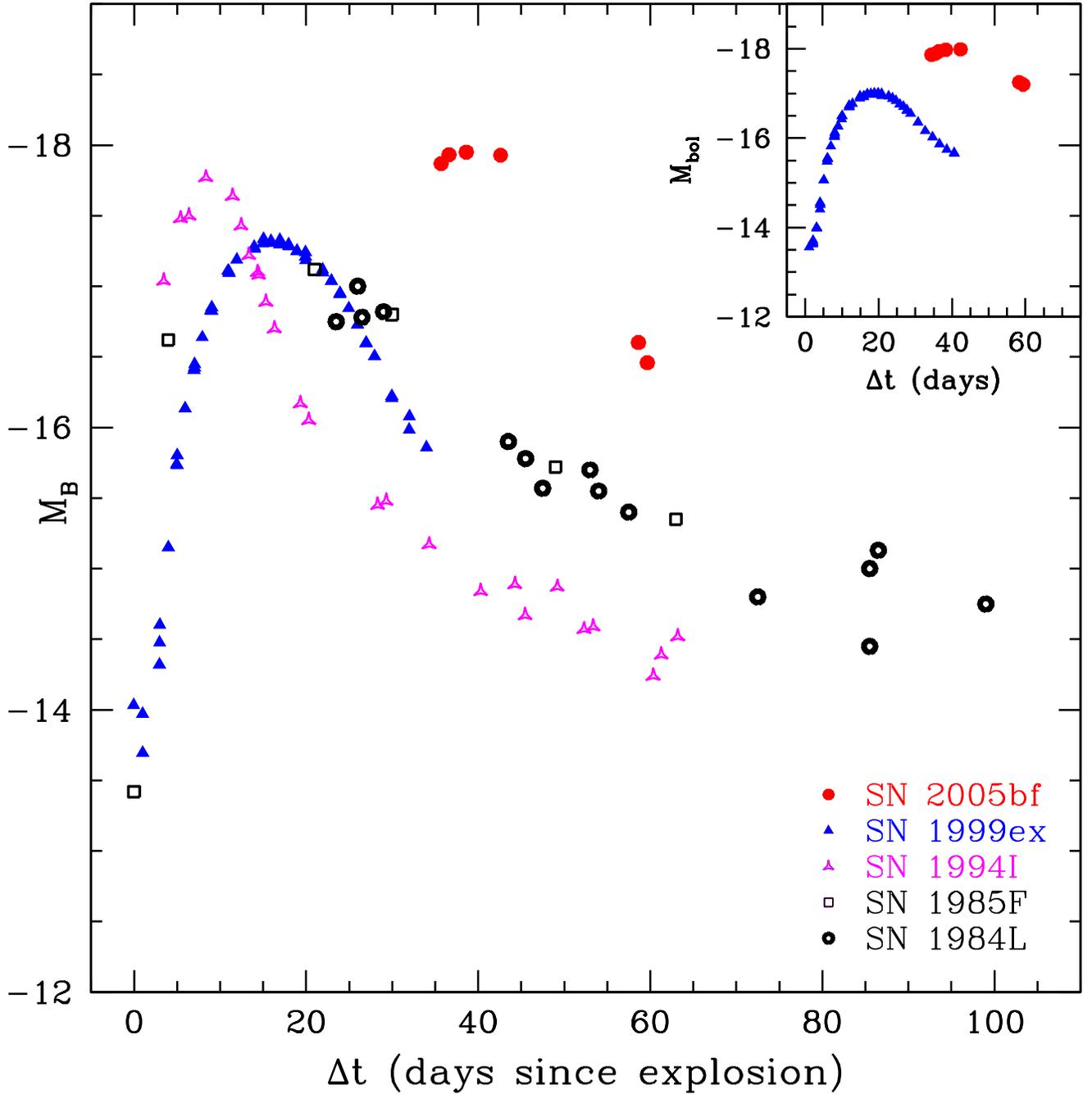}}
\caption{The evolution of $M_B$ of SN 2005bf from the day of explosion
compared with other type Ib/c SNe. Inset shows the bolometric light curve for
SN 2005bf (filled circles) and SN 1999ex (filled triangles) for a comparison.
\label{fig2}}
\end{figure}

The bolometric magnitudes were estimated by converting our $BVRI$ photometry,
corrected for the assumed E(B-V), into absolute monochromatic fluxes
adopting the magnitude-to-flux conversion factors compiled by \citet{bes98}.
The fluxes were then integrated using a fitting spline curve.
Around the light maximum, extending the spline fit only to 3600 \AA\ give
bolometric magnitudes about 0.15-0.2 mag fainter than if the fit were extended
to 3000 \AA, while
there is no significant difference around the epochs of our last observations,
indicating a significant contribution by the $U$ flux around maximum. Hence,
the bolometric magnitudes are estimated with zero-flux terminals of the
spline fit chosen as 3000 \AA\ and 2.480 $\mu$m in an effort to
recover as much as possible the $U$ and near-infrared fluxes that were missed
by our photometry. Adding a conservative uncertainty, $\pm0.2$, to the
bolometric magnitudes, we estimate the bolometric magnitude at maximum to be
$M_{\rm{bol}}=-18.0\pm 0.2$ on May 11 (JD\,2\,45\,3502.1).
We plot in Figure \ref{fig2} the evolution of the absolute magnitude in $B$ ($M_B$) for
SN 2005bf since our assumed date of explosion and compare
it with other SNe Ib/c, namely, SN 1999ex \citep{set02}, SN 1994I \citep{ret96},
SN 1984L \citep{t87, sk89} and SN 1985F \citep{t86}. The maximum bolometric
magnitude (inset in Figure \ref{fig2}) of SN 2005bf is brighter than the average value
for type Ib/c SNe, even though the time of maximum was significantly later.
Furthermore, the $B-V=0.37$ colour at maximum indicates SN 2005bf to be
marginally bluer.

The rise time depends on the mass and the explosion energy \citep[e.g.][]{nom04}.
The slow rise suggests a relatively low ratio of explosion energy to ejected
mass.  A detailed modelling of the light curve and the spectra are beyond
the scope of this work and will be reported in a later paper \citep{tom05}.
However, preliminary calculations indicate a tentative value for the
explosion energy of $\sim (1.0 - 1.5) 10^{51}$ erg, and an ejected mass of
$\sim 6 - 7 M_\odot$.  The brightness of the peak and its late
occurrence suggest a relatively large production of $^{56}$Ni ($\sim
0.31 M_\odot$), which points to a rather massive progenitor
($\sim 25 - 30 M_\odot$) \citep{tom05}.

\section{The Spectra}

\begin{figure}
{\includegraphics[width=18cm]{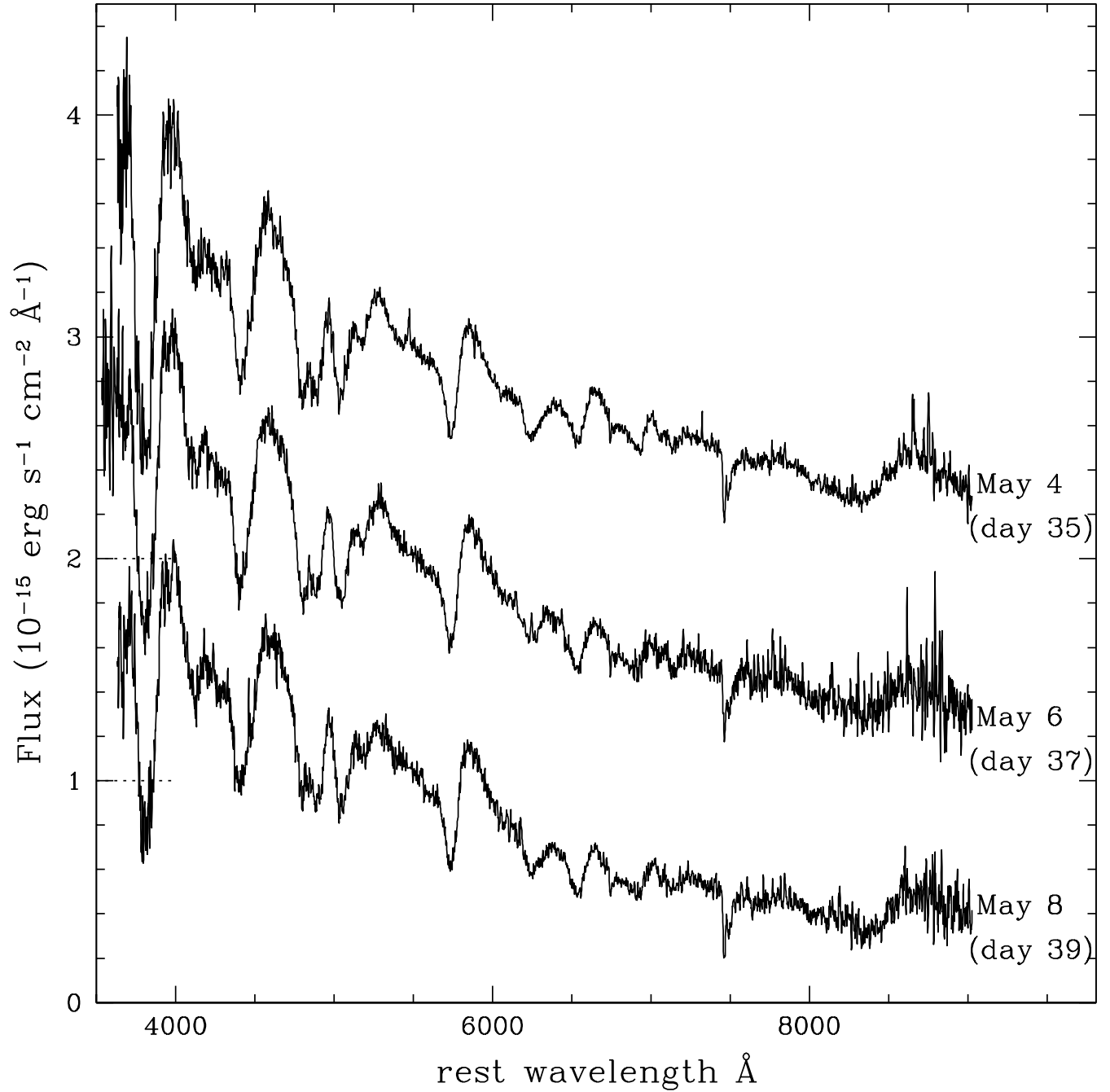}}
\caption{Optical spectra of SN 2005bf. The spectra are corrected for the
host galaxy redshift. Time (in days) since the date of explosion is indicated
for each spectrum. For clarity, the spectra have been displaced vertically.
Dotted lines at the left indicate the zero flux level for each spectrum. For
day 39, zero flux is the $x$-axis.
\label{fig3}}
\end{figure}

Spectra of SN 2005bf were obtained at a resolution of 8~\AA\ in the wavelength
range 3600--7200 \AA\ and 5200--9200 \AA\ on May 4.65, 6.62, 8.63 (UT) (marked
by vertical lines in Fig. \ref{fig1}). All
observations were made using a slit of 2.2 arcsec width and aligned along the
parallactic angle. Spectrophotometric standards HZ 44 and BD +33$^0$ 2642
observed on 2005 May 4 were used to correct the supernova spectra
for the response curves of the instrument and bring them to a flux scale. The
spectra in the two different regions were combined, scaled to a weighted mean,
to give the final spectrum on a relative flux scale, which were then
brought to an absolute scale using the $BVRI$ magnitudes. The flux calibrated
spectra, corrected for the redshift of the host galaxy are shown in Figure
\ref{fig3}.  The three spectra presented here are all near optical maximum, but
are very similar to those of SN 1984L \citep{har87, mat01} about one week past
maximum and SN 1999ex \citep{ham02} 4 days past maximum. If the phase since the
date of explosion is considered, then the spectra of SN 2005bf correspond to
about 35-39 days after explosion, while the corresponding phase is about 20-25
days for SN 1984L and SN 1999ex.

The spectra show prominent and broad P Cygni lines of He I, Fe II, and
Ca II. The He I $\lambda$5876 P Cygni feature is strong, whose
identification is supported by the presence of clear He I
$\lambda$6678 and $\lambda$7065, although Na I D could also
contribute. He I $\lambda$7281 may also exist. However, it should
be noted that this feature is affected by the telluric H$_2$O
absorption, and our spectra are not corrected for the telluric
features. The velocities corresponding to the absorption minima
of the relatively weak He I $\lambda$6678, $\lambda$7065, and $\lambda$7281
(if real) have average velocities of $\lesssim$ 6500 km~s$^{-1}$, lower than
that of He I $\lambda$5876 which is $\sim$ 7300 km~s$^{-1}$.
The very strong P Cygni feature between 3700 \AA\ and
4100 \AA\ and the very broad one between 8000 \AA\ and 9000 \AA\
are obviously Ca II H\&K and the near-infrared triplet,
respectively, with velocities $\gtrsim$ 10000 km~s$^{-1}$, indicating that
these lines have large optical depths. Between 4000 \AA\ and 5500 \AA\, the
spectra are dominated by Fe II multiplets, whose individual identifications
are difficult due to the large intrinsic number of Fe II optical
transitions and strong line-blending in the fast-moving SN
atmosphere. Nevertheless, we identify Fe II multiplet 27 ($\lambda$4233),
42 ($\lambda$4924, $\lambda$5018, and $\lambda$5169), and 49 ($\lambda$5317)
with velocities between $\sim$ 7500 km~s$^{-1}$ and $\sim$ 8000 km~s$^{-1}$.
The strong 4570\AA feature is a complex blend of several lines of Fe II
multiplets 37 and 38, mainly $\lambda\lambda$4629, 4584, 4549, and 4520. The
absorption velocity, calculated with respect to $\lambda$4520, a strong feature
in both multiplets, is consistent with other Fe II lines. The identity of the
P Cygni line between 6200 \AA\ and 6500 \AA\ is controversial. This feature,
if identified as H$\alpha$, corresponds to a velocity as high as
$\sim$ 15000 km~s$^{-1}$. If, instead, identified as Si II $\lambda$6355, the
measured velocity drops to $<5500$ km~s$^{-1}$, which is significantly lower
than all other lines. It may be noted that the uncertainty of our measurements
varies from line to line and from spectrum to spectrum and can be as large as
$\pm 500$ km~s$^{-1}$, a result of the low S/N ratio, potential weak lines, and
other pollution around the line absorption minima.

\begin{figure}
{\includegraphics[width=14cm]{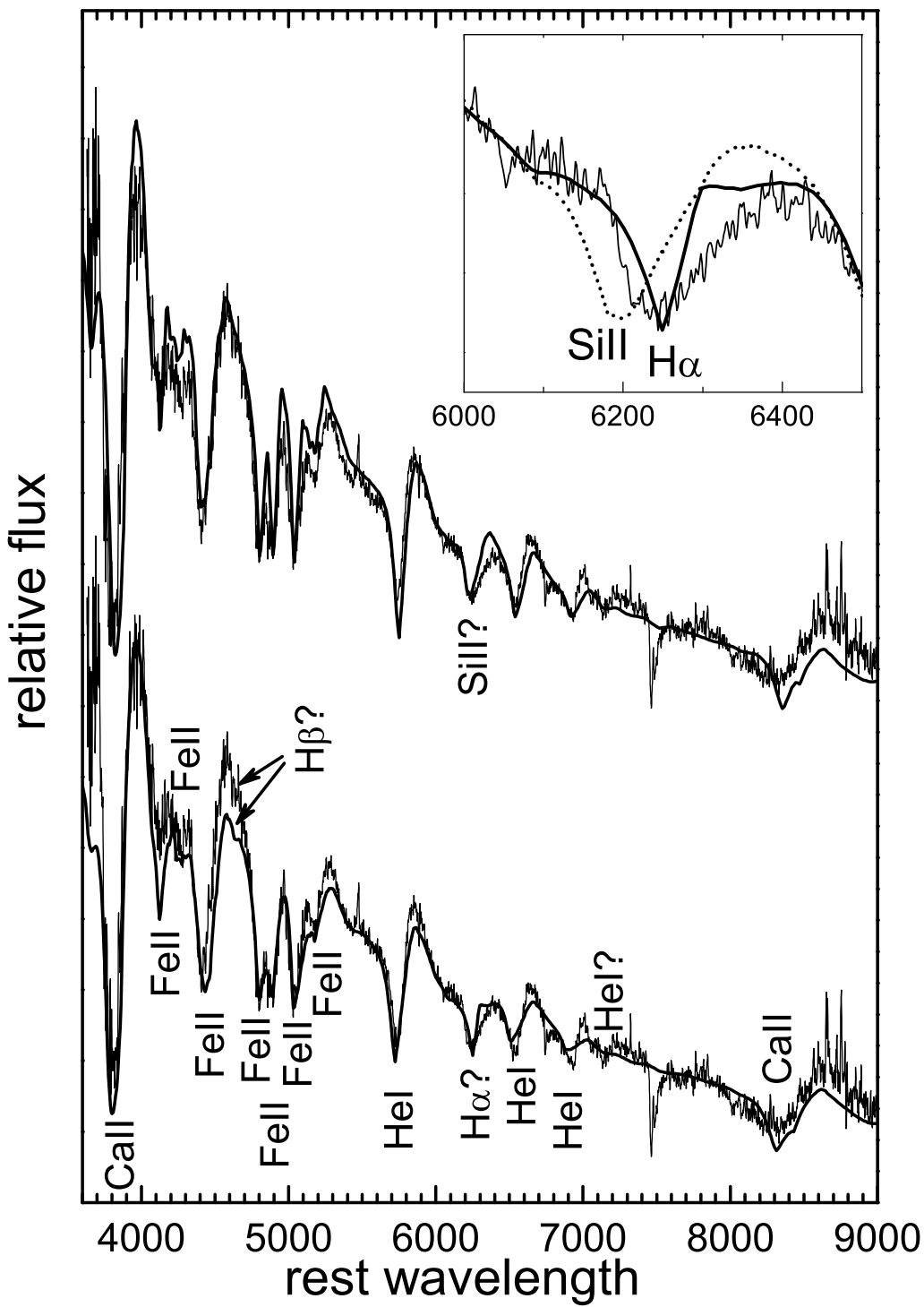}}
\caption{The day 35 (May 4) spectrum of SN 2005bf (thin line) compared
with a synthetic spectrum (lower spectrum thick line) that has
$v_{\rm{phot}}=8000$ km s$^{-1}$
and contains lines of He I, Ca II, Fe II and H. The thick line in the
upper spectrum is the synthetic spectrum without lines due to H, but Si II
included and $v_{\rm{phot}}=6500$ km s$^{-1}$. Inset shows the 6245 \AA\
absorption with fits due to H$\alpha$ (thick line) and Si II (dotted line).
\label{fig4}}
\end{figure}

To further establish line identifications, we compute synthetic
spectra using the fast, parameterized supernova spectrum-synthesis
code, SYNOW \citep[see][and references therein]{b02}, and
show the fit to the spectrum of May 4 in Figure \ref{fig4}. We assume a
$-7$ power law for the radial dependance of line optical depths.
The photospheric velocity, $V_{\rm ph}$, is assumed to be traced
by the absorption minima of weak lines.
We first assume $V_{\rm ph}=$ 8000 km~s$^{-1}$ ({\em lower thick solid line}),
the value that matches weak Fe II lines. As expected, Fe II and Ca II lines are
well reproduced, while He I $\lambda$6678 and $\lambda$7065 absorptions are a
bit bluer than the observed. The observed He I $\lambda$6678, $\lambda$7065,
and $\lambda$7281 (if real) are also stronger than in the model. This suggests
that non-thermal excitation, which is not included in SYNOW, is important for
He I. The $\sim$ 6240 \AA\ absorption minimum is reproduced by introducing a
high-velocity H$\alpha$ with a lower cut-velocity of 15,000 km s$^{-1}$
\citep[see also][]{wb05}. The narrow absorption and flat-topped emission of the
synthetic H$\alpha$ profile are consequences of the artifical
optical-depth discontinuity of a detached line. Identification of this
feature with Si II instead of H$\alpha$ produces too blue an absorption minimum
at 6200 \AA\ ({\em dotted line}, inset in Fig. 4). An alternative
identification of the feature is with Ne I $\lambda$6402 \citep{b03}.
A marginally
discernible dip at 4630 \AA\ in the Fe II peak may, if real, be explained by
high-velocity H$\beta$ (marked by {\em arrows} in Fig. 4). Hydrogen
lines have been suggested for other Ib SNe \citep[e.g.][]{den00, b03, whet94}.

We also computed a synthetic spectrum with $V_{\rm ph}=$ 6500 km~s$^{-1}$
({\em upper thick solid line}). This spectrum reproduces the positions of He I
absorption
minima, but Fe II absorptions are too red. As a possible solution, we
tentatively introduce a lower cut-velocity of 8000 km~s$^{-1}$ for Fe II.
One can assume Fe III dominates over Fe II below that velocity although Fe III
is actually not included in our spectrum synthesis. With such a low
$V_{\rm ph}$, Si II $\lambda$6355 seemingly matches the P Cygni feature
between 6200 \AA\ and 6500 \AA\ better than the high-$V_{\rm ph}$ case.
Calculations of realistic spatial structures of ionization and excitation
above the photosphere are needed to correctly identify this feature and to
determine the photospheric velocity, which is beyond the ability of SYNOW and
the scope of this paper.

\section{Conclusions}

The $BVRI$ light curve and spectra of SN 2005bf around maximum are presented.
The light curves indicate that the maximum occurred nearly 40 days after the
date of explosion. At maximum, SN 2005bf was brighter and bluer than other
SNe Ib/c. The maximum phase was broad and the decline rates slow,
and may be compared with the core collapse models of hydrogen-less cores.
Preliminary calculations suggest a core mass larger than the Type Ib model
suggested for SN1984L \citep{tom05}. The slow rise to the maximum
and the brighter peak bolometric luminosity indicate that most of $^{56}$Ni was
buried in a relatively low velocity region in the very massive ejecta
\citep{hac91}, although a small part of $^{56}$Ni may be mixed out 
\citep{tom05}.
The spectra of SN 2005bf around maximum are very similar to those of the Type
Ib SNe 1999ex and 1984L about 25-35 days after explosion with prominent He I,
Fe II, Ca II H\&K and the near-IR triplet P Cygni lines present. Relatively weak
H$\alpha$ and very weak H$\beta$ may also exist, blueshifted by $\sim 15,000$
km s$^{-1}$.
We suggest that SN 2005bf was the explosion of a massive He star, possibly with
a trace of hydrogen envelope.

\clearpage

\clearpage

\clearpage

\clearpage

\end{document}